\documentclass[12pt,aps,nofootinbib]{revtex4}
\usepackage{amsmath}
\usepackage{graphicx}

\def\be{\begin{equation}} \def\ee{\end{equation}} \def\bea{\begin{eqnarray}}
\def\eea{\end{eqnarray}} \def\nnb{\nonumber}

\begin{document}

\title{Effective Field Theory of $^3$He}
\author{Shung-ichi Ando}
\author{Michael C. Birse}
\affiliation{Theoretical Physics Group, School of Physics and Astronomy,
The University of Manchester, Manchester, M13 9PL, UK}

\begin{abstract}

$^3$He and the triton are studied as three-body bound states 
in the effective field theory without pions. We study $^3$He 
using the set of integral equations developed by Kok 
\textit{et al.} which includes the full off-shell 
$T$-matrix for the Coulomb interaction between the protons. 
To leading order, the theory contains: two-body contact 
interactions whose renormalized strengths are set by the $NN$ 
scattering lengths, the Coulomb potential, and a three-body 
contact interaction. We solve the three coupled integral equations 
with a sharp momentum cutoff, $\Lambda$, and find that a 
three-body interaction is required in $^3$He at leading order, 
as in the triton. It also exhibits the same limit-cycle behavior 
as a function of $\Lambda$, showing that the Efimov effect 
remains in the presence of the Coulomb interaction. We also
obtain the difference between the strengths of the three-body
forces in $^3$He and the triton.

\end{abstract}
\maketitle

\vskip 10pt

\section{Introduction}

Since Weinberg first proposed applying the ideas of effective field 
theory (EFT) to nuclear forces \cite{wein-90}, much effort has gone into 
this approach. (For reviews, see 
Refs.~\cite{betal-00,bk-arnps02,e-ppnp06}.) Although there is still some 
debate about how best to implement it at energies where pion-exchange 
forces are resolved, the picture is clearer at lower energies. Here 
few-nucleon systems can be described by a ``pionless" EFT based on two- 
and three-body contact interactions 
\cite{bvk-plb98,vk-npa99,ksw-plb98,bhk-npa00}. The 
resulting expansion of the two-body force is just that of the 
effective-range expansion \cite{b-pr49}, but the EFT framework makes it 
possible to extend this consistently to other effective operators and 
three-body forces.

This theory has been applied extensively to two-body systems, where
it has been extended to include the effects of the Coulomb interaction 
on proton-proton scattering 
\cite{kr-plb99,kr-prc01,bb-prc03,ashh-prc07,ab-prc08}. 
In that system, it corresponds to a distorted-wave or ``modified" version 
of the effective-range expansion \cite{b-pr49}.

Three-body systems have also been studied in the pionless EFT 
\cite{bvk-plb98,bhk-npa00}. In these, exchange of one particle between 
an interacting pair and the third particle leads to a long-range force 
that can be either attractive or repulsive depending on the overall
symmetry of the wave function. Attractive channels, such as the triton,
display the Efimov effect \cite{e-sjnp71} in the limit of infinite
two-body scattering length. This consists of an infinite tower of bound 
states with energies in a constant ratio. It results from a discrete
scale invariance in these systems and corresponds to a limit 
cycle of the renormalisation group (RG) for the contact three-body 
force \cite{bh-prl03}.

The $^3$He nucleus is of particular interest since it can potentially 
provide access to properties of the neutron that require a polarised 
target. It should be possible to describe its low-energy properties 
within the pionless EFT, extended to include the Coulomb interaction. 
However the only previous application of the theory to the $pd$ system 
is in the work of Rupak and Kong, who treated Coulomb effects perturbatively 
and only considered the spin-3/2 channel \cite{rk-npa03}. 

A nonperturbative treatment may not be essential for the $^3$He ground 
state, given the typical momenta in the wave function. However it will 
be necessary if the EFT is to be extended to describe $pd$ scattering near 
threshold. Also, it is needed to answer the question of whether the
Efimov effect survives in the presence of the Coulomb interaction.
Arguments based on the degrees of the singularities in the potentials
suggest it should and this has been checked by Hammer and Higa 
for a simpler two-body model \cite{hh-epja08}. However it has not 
previously been confirmed in a three-body system.

Like most other EFT studies of the triton and other three-body 
systems \cite{bk-arnps02}, our treatment is formulated in terms 
of a set of integral equations. However a quite different approach,
based on the resonating-group method, was recently applied to three- 
and four-nucleon systems by Kirscher \textit{et al.}~\cite{kgsh-epja10}.
While this also treats the Coulomb interaction to all orders, 
limitations on the range of cutoffs mean that it is not able to test 
whether the Efimov effect occurs.

Here we report the results of a study of $^3$He in the pionless EFT 
with the Coulomb interaction treated nonperturbatively. Our approach
is based on that used by Kok \textit{et al.} \cite{ksh-75} to study 
$^3$He with separable potentials and the full off-shell Coulomb 
$T$-matrix. The resulting set of integral equations is very similar to 
that developed by Skornyakov and ter-Martirosian\cite{stm-jetp57}
which have more recently been used in EFT studies of the triton and 
other three-body systems \cite{bk-arnps02}.

Our results show that the Efimov effect does indeed occur in the 
presence of the Coulomb interaction. We also determine the strength 
of the three-body force needed to reproduce observed $^3$He binding 
energy. The difference between this and the corresponding force 
needed for the triton is of the expected size for an electromagnetic 
effect.

\section{Lagrangian}

In the pionless EFT, the strong forces between nucleons are
described by two- and three-body contact interactions. In the present 
context it is convenient to represent the two-body interactions in terms 
of dibaryon fields. The resulting effective Lagrangian for the 
three-nucleon system can be written as \cite{bhk-npa00,ah-prc05}
\bea
{\cal L} &=& {\cal L}_N 
+{\cal L}_s
+{\cal L}_t
+{\cal L}_{3}\,,  
\eea
where ${\cal L}_N$ is the standard one-nucleon Lagrangian
in the heavy-baryon formalism,
\bea
{\cal L}_N &=&
N^\dagger \left\{
iv\cdot D 
+ \frac{1}{2m_N}\left[
(v\cdot D)^2 
-D^2\right]
\right\} N\,.
\eea
Here $v^\mu$ is a velocity vector satisfying a condition $v^2=1$, 
$D_\mu$ is the covariant derivative, and $m_N$ is the nucleon mass.
The terms ${\cal L}_s$ and ${\cal L}_t$ are dibaryon effective
Lagrangian for spin singlet and triplet parts, respectively,
and these read
\bea
{\cal L}_s &=&  s_a^\dagger \left\{
iv\cdot D
+ \Delta_{s(a)}
\right\}s_a
-y_s \left\{
s_a^\dagger \left[ 
N^T P_a^{({}^1S_0)}N
\right] + h.c.
\right\}\,,
\\ 
{\cal L}_t &=&  
 t_i^\dagger \left\{
iv\cdot D
+ \Delta_t
\right\}t_i
-y_t \left\{
t_i^\dagger \left[ 
N^T P_i^{({}^3S_1)}N
\right] + h.c.
\right\}\,,
\eea
where $s_a$ and $t_i$ are the corresponding dibaryon fields.
The strengths of the corresponding two-body interactions depend on
$\Delta_{s(a)}$ and $\Delta_t$, the mass differences between 
the dibaryons and two nucleons, and $y_s$ and $y_t$, the coupling 
constants for the dibaryon-nucleon-nucleon vertices. 
The projection operators for the two-nucleon ${}^1S_0$ and ${}^3S_1$ 
states are
\bea
P^{({}^1S_0)}_a = \frac{1}{\sqrt8} \tau_2\tau_a \sigma_2\,,
\ \ \ 
P^{({}^3S_1)}_i = \frac{1}{\sqrt8} \tau_2\sigma_2\sigma_i\,,
\eea
where $\tau_a$ and $\sigma_i$ are Pauli matrices for 
isospin and spin, respectively.

The present work is based on the leading-order (LO) terms in the expansion
of this EFT of $Q$. Here $Q$ stands for any of the low-energy scales in the 
problem: the relative momenta, $1/a$ where $a$ is any of the NN scattering 
lengths, and the inverse of the Bohr radius 
$\kappa=\alpha_{\scriptscriptstyle{\rm EM}} m_N/2$. 
More precisely, we keep all terms of order $Q^{-1}$, noting that the strong 
attraction in the NN $S$ waves enhances the wave functions near the origin
and hence promotes contact interactions proportional to $1/a$ or $\kappa$ to 
this order \cite{vk-npa99,ksw-plb98,bmr-plb99,b-pos09}.

The Coulomb interaction leads to a well-known logarithmic divergence in the 
NN loop integrals \cite{kr-plb99,kr-prc01,bb-prc03}. This is renormalised by a LO 
counterterm that contributes to the $pp$ scattering length. Given that we need to 
include isospin breaking in this channel, we have also taken account of
charge-independence breaking between the $^1S_0$ $np$ and $nn$ scattering lengths,
even though this is formally of higher order. This allows us to present the general
structures of the three-body equations, which will be needed for extensions of this 
work to higher orders. We have not included the neutron-proton mass difference 
which is not only of higher order but is smaller than might be expected
as a result of cancellations between electromagnetic and strong contributions
\cite{gl-pr82}. To the order we work, observables depend only on the combinations 
$\Delta_{s(a)}/y_s^2$ and so we have chosen to leave the couplings $y_s$ 
isospin-invariant.

The three-body force is expressed as a dibaryon-nucleon contact
interaction. It is given by the effective Lagrangian,
\bea
{\cal L}_3 &=& \frac{m_N H(\Lambda)}{3\Lambda^2}\,\Bigl\{
y_t^2 N^\dagger (
\vec{\sigma}\cdot\vec{t})^\dagger (\vec{\sigma}\cdot\vec{t})N
-y_sy_t \left[
N^\dagger (\vec{\sigma}\cdot\vec{t})^\dagger (\vec{\tau}\cdot\vec{s})N +
h.c. \right] 
\nnb \\ && 
\qquad\qquad\qquad+\, y_s^2 N^\dagger (\vec{\tau}\cdot\vec{s})^\dagger 
(\vec{\tau}\cdot\vec{s})N \Bigr\}\,,
\label{eq;L3}
\eea
where $H(\Lambda)$ is the coupling constant, which runs with the scale 
$\Lambda$ of the cutoff we impose on the coupled integral equations.
Note that this interaction contributes only to three-body channels
with total spin 1/2 and total isospin 1/2. 

The building blocks needed to construct the three-body integral 
equations from the Lagrangian are as follows. In the one-body sector, 
we have the non-relativistic nucleon propagator,
\bea
iS_N(p) &=& \frac{i}{p_0 - \frac{\vec{p}^2}{2m_N}+i\epsilon}\,.
\eea
In the two-body sector, we need the dressed dibaryon propagators with and 
without the Coulomb interaction.
\begin{figure}[h]
\includegraphics[width=15cm,keepaspectratio,clip]{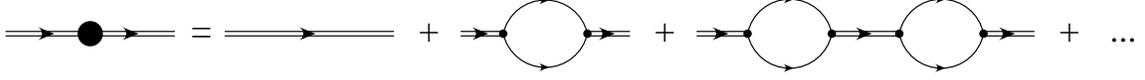}
\caption{Diagrams for the dressed dibaryon propagator, denoted by the double line
with filled circle. Single lines represent nucleon propagators; double lines 
undressed dibaryon propagators.}
\label{fig;dibaryonpropagator}
\end{figure}
The propagators for the $np$ and $nn$ channels,
which have no Coulomb interaction, are represented by the
diagrams in Fig.~\ref{fig;dibaryonpropagator} and are given by
\bea
iD_{s,t}\left(E-\frac{3q^2}{4m_N}\right) &=& 
\frac{4\pi}{m_Ny^2_{s,t}}\;
\frac{-i}{\,-\mu -\frac{4\pi \Delta_{s,t}}{m_N y_{s,t}^2}
+ \sqrt{\frac34q^2 -m_NE-i\epsilon}
+i\epsilon}\,.
\eea

The two-nucleon loop diagrams here have been dimensionally regularised 
using the PDS scheme with the subtraction scale $\mu$. The dependence 
on this can be removed by renormalising the constants 
$4\pi \Delta_{s,t}/(m_Ny_{s,t}^2)$ using
\bea
\mu+\frac{4\pi\Delta_{s(np)}}{m_Ny_s^2} = \frac{1}{a_{np}}\,,
\ \ \ 
\mu+\frac{4\pi\Delta_{s(nn)}}{m_Ny_s^2} = \frac{1}{a_{nn}}\,,
\ \ \ 
\mu+\frac{4\pi\Delta_t}{m_Ny_t^2} = \gamma\,,
\eea
where $a_{np}$ and $a_{nn}$ are the scattering lengths 
for the spin-singlet $np$ and $nn$ channels, respectively,
and $\gamma$ is related to the deuteron binding energy
$B_2$ through $\gamma=\sqrt{m_NB_2}$. Working to leading order, 
we neglect corrections from the effective ranges.

\begin{figure}[h]
\includegraphics[width=15cm,keepaspectratio,clip]{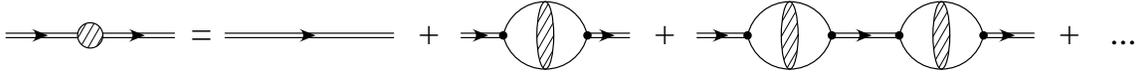}
\caption{Diagrams for dressed dibaryon propagator for the $pp$
channel, denoted by the double line with shaded circle. The shaded ovals 
here represent the two-nucleon Green's function
dressed with the Coulomb interaction.}
\label{fig;dibaryonpropagator-with-coulomb}
\end{figure}

In the $pp$ channel, the Coulomb interaction dresses the two-nucleon
Green's function in the bubble diagram for the dressed dibaryon propagator, 
as shown in Fig.~\ref{fig;dibaryonpropagator-with-coulomb}. The resulting
propagator is given by 
\bea
iD_{s(pp)}\left(E-\frac{3q^2}{4m_N}\right) &=& \frac{4\pi}{m_Ny_s^2}\;
\frac{-i}{\,-\,\frac{4\pi\Delta_{s(pp)}^R}{my_s^2} 
- 2\kappa H(\kappa/p')}\,,
\eea
where 
\bea
H(\eta) &=& \psi(i\eta) 
+\frac{1}{2i\eta} 
-\ln(i\eta)\,, 
\eea
$\psi$ is the logarithmic derivative of the $\Gamma$ function,
and $-ip'=\sqrt{\frac34q^2-m_NE-i\epsilon}$.
The renormalised constant $\Delta_{s(pp)}^R$ for the $pp$ channel 
cancels both the linear and logarithmic divergences of the loop
diagram. It is related to the Coulomb scattering length $a_C$ by
\bea
\frac{1}{a_C}&=&\frac{4\pi \Delta_{s(pp)}^R}{m_Ny_s^2} 
=\frac{4\pi \Delta_{s(pp)}}{m_Ny_s^2} +\mu
-2\kappa\left[1-C_E+\ln\left(\frac{\mu}{4\kappa}\right)\right]
\,,
\eea
where $C_E=0.577215\cdots$ is Euler's constant. Note that the logarithmic 
divergence means that it is impossible to make a model-independent 
decomposition of $1/a_C$ into strong and electromagnetic contributions 
\cite{kr-plb99,kr-prc01,bb-prc03}. This implies that within this EFT there 
is no unambiguous way to separate Coulomb from other isospin-breaking
effects.

The final building block for the integral equations is the off-shell 
Coulomb T-matrix. A convenient form for it is the integral
representation \cite{cc-72} for negative energies, $k^2/m_N<0$: 
\bea
\langle {\bf p}'|T_C(k^2/m_N)|{\bf p}\rangle &=&
\frac{e^2}{({\bf p}'-{\bf p})^2}\left[
1-4i\eta\int^1_0 dt\, \frac{t^{i\eta}}{4t-(1-t)^2(x^2-1)}
\right]\,,
\eea
where $\eta =\kappa/k$ and 
\bea
x^2 = 1 + \frac{(p'^2-k^2)(p^2-k^2) }{k^2({\bf p}'-{\bf p})^2}
\,.
\eea

\section{Integral equations}

Now we construct the integral equations for scattering of a third nucleon
off a deuteron, concentrating on the channels where the third
nucleon is in an $S$ wave, and the total spin and isospin of the three
particles are both 1/2. The negative-energy solutions of these describe 
the bound triton and $^3$He states. 

We work the center of mass frame for the $Nd$ scattering and use the 
notation of Bedaque \textit{et al.}~\cite{bhk-npa00}. In the 
isospin-symmetric case, this process can be described by an amplitude 
$a(p,k)$ for scattering into states with a $^3S_1$ dibaryon, and 
$b(p,k)$ for scattering into those with a $^1S_0$ dibaryon. 
Here $k$ denotes the initial (on-shell) relative momentum of the 
deuteron and the third nucleon, and $p$ the final (off-shell) momentum.
 
\begin{figure}[h]
\includegraphics[width=14cm,keepaspectratio,clip]{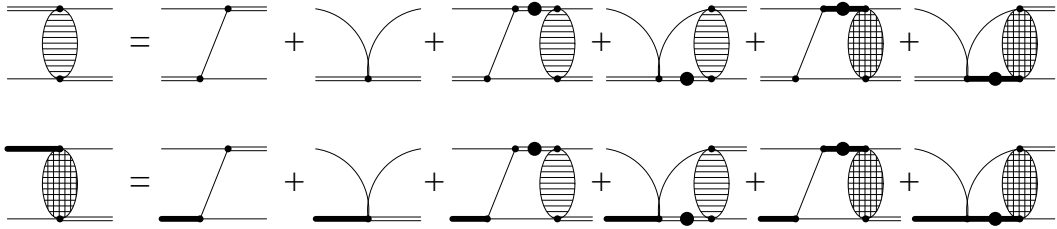}
\caption{Diagrams corresponding to the coupled integral equations for
nucleon-dibaryon scattering. Thin lines denote nucleons, double and 
thick lines $^3S_1$ and $^1S_0$ dibaryons. Those with filled circles 
are dressed dibaryon propagators, as in Fig. \ref{fig;dibaryonpropagator}.
Ovals with stripes denote the off-shell amplitude $a(p,k)$
with initial and final spin-triplet dibaryons, and those with cross 
stripes the amplitude $b(p,k)$ with initial spin-triplet and final 
spin-singlet dibaryon fields.}
\label{fig;integral-equations}
\end{figure}

The amplitudes satisfy coupled integral equations which can be 
represented diagrammatically as in Fig.~\ref{fig;integral-equations}.
The kernels of these equations consist of two terms: one-nucleon-exchange,
which provides a long-range force between the dibaryon and the third
nucleon, and the three-body contact interaction. 

In the limit of infinite two-body scattering lengths (and zero deuteron
binding energy) the long-range force is scale-independent. It is this feature
that leads to the Efimov effect in channels where the force is attractive.

When we include the Coulomb interaction, isospin symmetry is broken
and we must introduce separate amplitudes $b_+(p,k)$ and $b_0(p,k)$ 
for $pd$ scattering into states with $^1S_0$ $pp$ and $np$ 
dibaryons, respectively. These and $a(p,k)$ satisfy a set of three coupled 
integral equations. For completeness, we also present the corresponding
set of three equations that arise when we allow for isospin-breaking effects 
in the $nd$ case, involving the amplitude $b_-(p,k)$ for states with $nn$ 
dibaryons. As already mentioned, these are not essential for the present LO
calculations but will be needed for extensions to higher orders.

\subsection{Integral equations for the $nd$ channel}

Before presenting the equations for $pd$ scattering, we first
look at the simpler equations for $nd$ scattering. Allowing for 
isospin breaking in the scattering lengths, the three integral 
equations for the $S=1/2$, $T=1/2$ channel are
\bea
a(p,k) &=& 
K^{(a)}(p,k;E) + 2\, \frac{H(\Lambda)}{\Lambda^2}
\nnb \\ && 
+\, \frac{1}{\pi}\int^\Lambda_0 dq\,q^2 \left[
K^{(a)}(p,q;E) + 2\, \frac{H(\Lambda)}{\Lambda^2}
\right] \frac{a(q,k)}{
-\gamma 
+ \sqrt{\frac34q^2 -mE}}
\nnb \\ && 
+\, \frac{1}{\pi}\int^\Lambda_0 dq\,q^2 \left[
 K^{(a)}(p,q;E) 
+ \frac23\, \frac{H(\Lambda)}{\Lambda^2}
\right] \frac{b_0(q,k)}{
-\frac{1}{a_{np}}
+ \sqrt{\frac34q^2 -mE}}
\nnb \\ && 
+\, \frac{2}{\pi}\int^\Lambda_0 dq\,q^2 \left[
K^{(a)}(p,q;E) 
+ \frac23\, \frac{H(\Lambda)}{\Lambda^2}
\right] \frac{b_-(q,k)}{
-\frac{1}{a_{nn}}
+ \sqrt{\frac34q^2 -mE}}\,,
\label{eq;a-iso-b}
\\
b_0(p,k) &=& 3 K^{(a)}(p,k;E) + 2\, \frac{H(\Lambda)}{\Lambda^2}
\nnb \\ && 
+\, \frac{1}{\pi}\int^\Lambda_0 dq\,q^2 \left[
3 K^{(a)}(p,q;E) 
+ 2\, \frac{H(\Lambda)}{\Lambda^2}
\right] \frac{a(q,k)}{
-\gamma 
+ \sqrt{\frac34q^2 -mE}}
\nnb \\ && 
+\, \frac{1}{\pi}\int^\Lambda_0 dq\,q^2 \left[
-K^{(a)}(p,q;E) 
+ \frac23\, \frac{H(\Lambda)}{\Lambda^2}
\right] \frac{b_0(q,k)}{
-\frac{1}{a_{np}}
+ \sqrt{\frac34q^2 -mE}}
\nnb \\ && 
+\, \frac{2}{\pi}\int^\Lambda_0 dq\,q^2 \left[
K^{(a)}(p,q;E) 
+ \frac23\, \frac{H(\Lambda)}{\Lambda^2}
\right] \frac{b_-(q,k)}{
-\frac{1}{a_{nn}}
+ \sqrt{\frac34q^2 -mE}}\,,
\label{eq;b0-iso-b}
\\
b_-(p,k) &=& 3 K^{(a)}(p,k;E) + 2\, \frac{H(\Lambda)}{\Lambda^2}
\nnb \\ && 
+\, \frac{1}{\pi}\int^\Lambda_0 dq\,q^2 \left[
3 K^{(a)}(p,q;E) 
+ 2\, \frac{H(\Lambda)}{\Lambda^2}
\right] \frac{a(q,k)}{
-\gamma 
+ \sqrt{\frac34q^2 -mE}}
\nnb \\ && 
+\, \frac{1}{\pi}\int^\Lambda_0 dq\,q^2 \left[
K^{(a)}(p,q;E) 
+ \frac23\, \frac{H(\Lambda)}{\Lambda^2}
\right] \frac{b_0(q,k)}{
-\frac{1}{a_{np}}
+ \sqrt{\frac34q^2 -mE}} 
\nnb \\ && 
+\, \frac{2}{\pi}\int^\Lambda_0 dq\,q^2 \left[
\frac23\, \frac{H(\Lambda)}{\Lambda^2}
\right] \frac{b_-(q,k)}{
-\frac{1}{a_{nn}}
+ \sqrt{\frac34q^2 -mE}}\,,
\label{eq;bm-iso-b}
\eea
where
\bea
K^{(a)}(p,q;E) &=&
\frac{1}{2pq}\ln\left(
\frac{p^2+q^2+2pq-m_NE}{
p^2+q^2-2pq-m_NE}
\right)\,.
\label{eq;one-kernel}
\eea
If  we set $a_{nn}=a_{np}$ and $b_0=b_-=b$, these reduce to the 
isospin-symmetric forms of the equations in Ref.~\cite{bhk-npa00}. 

The integrals over the relative momentum $q$ are all cut off at
$q=\Lambda$. The resulting dependence on $\Lambda$ can be cancelled 
by the three-body force, whose strength is $H(\Lambda)/\Lambda^2$.

\subsection{Integral equations for the $pd$ channel}

We now turn to the corresponding equations for $pd$ scattering.
These differ from the ones in the previous subsection by having
$b_+$ instead of $b_-$ and, more importantly, adding the 
Coulomb interaction between the two protons. They can be obtained 
from the equations developed by the Groningen group \cite{ksh-75}, 
by omitting the separable form factors and instead
imposing a sharp cutoff on the relative momenta.

\begin{figure} [h]
\includegraphics[width=12cm,keepaspectratio,clip]{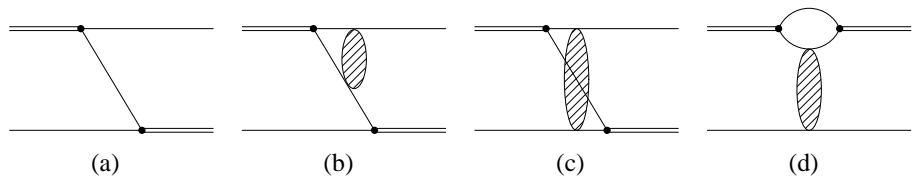}
\caption{Diagrams corresponding to the long-range terms in the kernel
for $pd$ scattering. Here, the shaded oval denotes the off-shell Coulomb 
$T$-matrix (since contributions where the two protons do not interact are already
contained in (a)). Diagram (b) has been shown for the case of an initial $np$ 
dibaryon and a final $pp$ one. The corresponding diagram for the inverse 
process is the mirror image of this.}
\label{fig;Fp}
\end{figure}

The Coulomb interaction leads to additional long-range terms in the 
kernels of these equations. The diagrams for the full long-range kernel 
are shown in Fig.~\ref{fig;Fp}. The one-nucleon exchange 
term of diagram (a) is supplemented by diagram (b) if a proton is 
exchanged, and by (c) for neutron exchange. Finally, diagram (d) 
represents the Coulomb interaction between an $np$ dibaryon and 
a proton. 

The coupled integral equations can be written
\bea
a(p,k) &=& 
K^{(a)}(p,k;E) 
+ K^{(c)}(p,k;E)
+ 2 K^{(d)}(p,k;E)
+ 2\, \frac{H(\Lambda)}{\Lambda^2} 
\nnb \\ && 
+\, \frac{1}{\pi}\int^\Lambda_0 dq\,q^2 \left[
K^{(a)}(p,q;E) 
+ K^{(c)}(p,k;E)
+ 2 K^{(d)}(p,k;E)
+ 2\, \frac{H(\Lambda)}{\Lambda^2}
\right] \nnb \\ && 
\qquad\qquad\qquad\times\,\frac{a(q,k)}{-\gamma 
+ \sqrt{\frac34q^2 -mE}}
\nnb \\ && 
+\, \frac{1}{\pi}\int^\Lambda_0 dq\,q^2 \left[
K^{(a)}(p,q;E) 
+  K^{(c)}(p,k;E)
+ \frac23\, \frac{H(\Lambda)}{\Lambda^2}
\right] \frac{b_0(q,k)}{
-\frac{1}{a_{np}}
+ \sqrt{\frac34q^2 -mE}}
\nnb \\ && 
+\, \frac{2}{\pi}\int^\Lambda_0 dq\,q^2 \left[
 K^{(a)}(p,k;E) 
+  K^{(b)}_{13}(p,k;E)
+ \frac23\, \frac{H(\Lambda)}{\Lambda^2}
\right] \frac{b_+(q,k)}{
-\frac{1}{a_{C}}
-2\kappa H(\kappa/p')}\,,\nnb \\ &&
\\
b_0(p,k) &=& 
 3 K^{(a)}(p,k;E)
+ 3 K^{(c)}(p,k;E)
+ 2\, \frac{H(\Lambda)}{\Lambda^2}
\nnb \\ && 
+\, \frac{1}{\pi}\int^\Lambda_0 dq\,q^2 \left[
3 K^{(a)}(p,q;E) 
+ 3 K^{(c)}(p,q;E) 
+ 2\, \frac{H(\Lambda)}{\Lambda^2}
\right] \frac{a(q,k)}{
-\gamma 
+ \sqrt{\frac34q^2 -mE}}
\nnb \\ && 
-\, \frac{1}{\pi}\int^\Lambda_0 dq\,q^2 \left[
K^{(a)}(p,q;E) 
+ K^{(c)}(p,q;E) 
+ 2 K^{(d)}(p,q;E) 
- \frac23\, \frac{H(\Lambda)}{\Lambda^2}
\right]\nnb \\ && 
\qquad\qquad\qquad\times\, \frac{b_0(q,k)}{
-\frac{1}{a_{np}}
+ \sqrt{\frac34q^2 -mE}}
\nnb \\ && 
+\, \frac{2}{\pi}\int^\Lambda_0 dq\,q^2 \left[
K^{(a)}(p,q;E) 
+ K^{(b)}_{13}(p,q;E) 
+ \frac23\, \frac{H(\Lambda)}{\Lambda^2}
\right] \frac{b_+(q,k)}{
-\frac{1}{a_{C}}
-2\kappa H(\kappa/p')}\,,\nnb \\ &&
\\
b_+(p,k) &=& 3 
 K^{(a)}(p,k;E) 
+ 3 K^{(b)}_{31}(p,k;E) 
+ 2\, \frac{H(\Lambda)}{\Lambda^2}
\nnb \\ && 
+\, \frac{1}{\pi}\int^\Lambda_0 dq\,q^2 \left[
3 K^{(a)}(p,q;E) 
+ 3 K^{(b)}_{31}(p,q;E) 
+ 2\, \frac{H(\Lambda)}{\Lambda^2}
\right] \frac{a(q,k)}{
-\gamma 
+ \sqrt{\frac34q^2 -mE}}
\nnb \\ && 
+\, \frac{1}{\pi}\int^\Lambda_0 dq\,q^2 \left[
K^{(a)}(p,k;E) 
+ K^{(b)}_{31}(p,q;E) 
+ \frac23\, \frac{H(\Lambda)}{\Lambda^2}
\right] \frac{b_0(q,k)}{
-\frac{1}{a_{np}}
+ \sqrt{\frac34q^2 -mE}} 
\nnb \\ && 
+\, \frac{2}{\pi}\int^\Lambda_0 dq\,q^2 \left[ 
\frac23\, \frac{H(\Lambda)}{\Lambda^2}
\right] \frac{b_+(q,k)}{
-\frac{1}{a_C}
-2\kappa H(\kappa/p')}\,,
\eea
where $K^{(a)}$ is the one-nucleon exchange kernel defined in
Eq.~(\ref{eq;one-kernel}) above, and $K^{(b)}_{31}$, 
$K^{(b)}_{13}$, $K^{(c)}$, and $K^{(d)}$ are the pieces corresponding 
to the Coulomb diagrams (b), (c), (d) in Fig.~\ref{fig;Fp}. 
(In $K_{31}^{(b)}$ and $K_{13}^{(b)}$, the subscripts 1 and 3
refer to initial and final dibaryons, 1 denoting $np$ 
and 3 $pp$.)

In diagram (b), the dibaryon-nucleon-nucleon vertex projects out
only the $S$-wave part of the Coulomb T-matrix. This considerably
simplifies the corresponding terms in the equations by
reducing the number variables that need to be integrated
from six to two. The corresponding terms in the kernels are
\bea
K_{13}^{(b)}(p,q;E) &=& 
\frac12\int^1_{-1}dy\,
\frac{\phi^{(b)}_{13}(p,q,y;E)}{p^2+q^2-mE+pqy}\,,
\\
K_{31}^{(b)}(p,q;E) &=& 
\frac12\int^1_{-1}dy\,
\frac{\phi^{(b)}_{31}(p,q,y;E)}{p^2+q^2-mE+pqy}\,,
\eea
where $y=\hat{p}\cdot\hat{q}=\cos\theta$ and
\bea
\phi_{13}^{(b)}(p,q,y;E) &=& 
- \frac{4\kappa\beta'}{p'^2+\beta'^2}\int^1_0 dt\,t^{\kappa/\beta'}
\frac{1}{1+t^2-2\left(\frac{p'^2-\beta'^2}{p'^2+\beta'^2}\right)t}\,,
\\
\phi_{31}^{(b)}(p,q,y;E) &=& 
- \frac{4\kappa\beta''}{p''^2+\beta''^2}\int^1_0 dt\,t^{\kappa/\beta''}
\frac{1}{1+t^2-2\left(\frac{p''^2-\beta''^2}{p''^2+\beta''^2}\right)t}\,.
\eea
Here we have defined the momentum variables 
\bea
p' = \left|{\bf p}+\frac12{\bf q}\right|\,, 
\qquad
p'' = \left|{\bf q}+\frac12{\bf p}\right|\,, 
\qquad
\beta' = \sqrt{\frac34q^2-mE}\,,
\qquad 
\beta'' = \sqrt{\frac34p^2-mE}\,.
\eea

The kernels for diagrams (c) and (d) involve six dimensional integrals but 
we can analytically evaluate two of these (over the azimuthal angles). 
In the case of $K^{(c)}$, it is convenient to introduce the vectors,
\bea
{\bf a} = {\bf p}+{\bf q}\,, 
\qquad
{\bf b} = {\bf p}-{\bf q}\,,
\eea 
Choosing axes where ${\bf a}$ lies along the $z$-direction,
these can be written
\bea
{\bf a} = a\,(0,0,1)\,,
\qquad
{\bf b} = b\,(\sin\theta'',0,\cos\theta'')\,,
\qquad
{\bf l} = l\,(\sin\theta'\cos\phi',\sin\theta'\sin\phi',\cos\theta')\,,
\eea
in terms of their magnitudes
\be
a = \sqrt{p^2+q^2+2pq\cos\theta}\,,
\qquad
b = \sqrt{p^2+q^2-2pq\cos\theta}\,,
\ee
and the angles defined by
\be
\sin\theta'' =
\frac{2pq\sin\theta}{\sqrt{(p^2-q^2)^2+4p^2q^2\sin^2\theta}}\,,
\qquad 
\cos\theta'' =
\frac{p^2-q^2}{\sqrt{(p^2-q^2)^2+4p^2q^2\sin^2\theta}}\,.
\ee
The kernel then becomes
\bea
K^{(c)}(p,q;E) &=&  -\frac{\kappa}{\pi}\int^\infty_0 dl
\int^1_{-1}d(\cos\theta)
\int^1_{-1}d(\cos\theta')
\int^1_0 dt\,\frac{(\kappa/\beta)t^{\kappa/\beta-1}}{
bl\sqrt{d^2+4t(1-t)^{-2}l^2\beta^2}}\nnb \\ 
&& \qquad\times\left(
\frac{1}{\cos\theta''\cos\theta'+\frac{2}{bl}\sqrt{
d^2+4t(1-t)^{-2}l^2\beta^2}}\,
\frac{1}{\sqrt{1-y_1^2}}
\right. \nnb \\ 
&& \qquad\qquad\left.
-\, \frac{1}{\cos\theta''\cos\theta'-\frac{2}{bl}
\sqrt{d^2+4t(1-t)^{-2}l^2\beta^2}}\,
\frac{1}{\sqrt{1-y_2^2}}
\right)\,,
\label{eq;Kc}
\eea
where
\bea
y_{1,2} &=& \frac{\sin\theta''\sin\theta'}{\cos\theta''\cos\theta'
\pm\frac{2}{bl}\sqrt{d^2+4t(1-t)^{-2}l^2\beta^2}}\,,
\nnb \\
d&=& l^2+p^2+q^2+pq\cos\theta-\frac32al\cos\theta'-m_NE\,,
\nnb \\
\beta^2 &=& \frac34(l^2+a^2+2al\cos\theta')-m_NE\,.
\eea

The last term in the kernel, arising from diagram (d), requires 
the most care in its numerical evaluation. This is because the 
long-range photon exchange gives rise to a logarithmic IR singularity.
In the present work, where we are interested only in bound states,
the finite extent of the wave functions helps to regulate this 
singularity. Calculations of $pd$ scattering amplitudes will, however,
require different numerical methods. Following Ref.~\cite{ksh-75},
we split $K^{(d)}$ up into three pieces,
\bea
K^{(d)}(p,q;E) &=&  
K^{(d,V')}(p,q;E) 
+K^{(d,V'')}(p,q;E) 
+K^{(d,T-V)}(p,q;E) \,,
\label{eq;Kd}
\eea
where $K^{(d,V')}$ and $K^{(d,V'')}$ 
denote the contributions the from singular and nonsingular parts
of one-photon exchange, and $K^{(d,T-V)}$ the remaining (also
nonsingular) terms from iterating the Coulomb potential.
These have pieces have the forms
\bea
K^{(d,V')}(p,q;E) &=& -\, \frac{\kappa}{4pq\beta'}\ln\left(
\frac{p^2+q^2+2pq}{
p^2+q^2-2pq}
\right)\,, 
\label{eq;KdVp}
\\
K^{(d,V'')}(p,q;E) &=&
-\kappa \int^1_{-1}d(\cos\theta)\,\frac{1}{p^2+q^2-2pq\cos\theta}\left\{
\frac{1}{\sqrt{p^2+q^2-2pq\cos\theta}}
\right.
\nnb \\
 &&  \times\left[\arctan\left(
\frac{-p^2+2q^2-pq\cot\theta}{2\beta'\sqrt{p^2+q^2-2pq\cos\theta}}
\right)\right.
\nnb \\
&&\qquad\left.\left.
+\arctan\left(
\frac{2p^2-q^2-pq\cos\theta}{2\beta\sqrt{p^2+q^2-2pq\cos\theta}}
\right)
\right]
-\frac{1}{2\beta'}
\right\}\,,
\\
K^{(d,T-V)}(p,q;E) &=& -\kappa\int^1_{-1}d(\cos\theta)\,
\frac{1}{p^2+q^2-2pq\cos\theta}
\nnb \\ && \times \Biggl\{
\frac{1}{\pi}\int^\infty_0dl\,l^2\int^1_{-1}d(\cos\theta')\int^1_0dt\,
\frac{(\kappa/\beta_l)t^{\kappa/\beta_l-1}}{D\sqrt{1-c^2}}
\nnb \\ &&
\qquad-\,\frac{1}{\sqrt{p^2+q^2-2pq\cos\theta}}\left[
\arctan\left(
\frac{-p^2+2q^2pq\cos\theta}{2\beta'\sqrt{p^2+q^2-2pq\cos\theta}}
\right)\right.
\nnb \\ &&\left.\left.
\qquad+\arctan\left(
\frac{2p^2-q^2-pq\cos\theta}{2\beta\sqrt{p^2+q^2-2pq\cos\theta}}
\right)
\right]\right\}\,,
\eea  
where
\bea
\beta' &=& \sqrt{\frac34p^2-m_NE}\,,
\ \ \ 
\beta = \sqrt{\frac34q^2-m_NE}\,,
\ \ \ 
\beta_l = \sqrt{\frac34l^2-m_NE}\,,
\nnb \\ 
D &=& (l^2+q^2+ql\cos\theta-m_NE)(l^2+p^2+pl\cos\theta-m_NE)
\nnb \\ && + 4t(1-t)^{-2}\beta_l^2 (p^2+q^2-2pq\cos\theta)\,,
\nnb \\
c &=& (l^2+q^2+ql\cos\theta-m_NE)pl\sin\theta\sin\theta'/D\,.
\eea

\section{Numerical results for bound states}

We solve for the three-body bound states by taking
the homogeneous parts of the coupled integral equations
and treating them as nonlinear eigenvalue problems,
\be
K(E)\,u= u \,,
\label{eq;Xx}
\ee
where $u^T=(a,b_0,b_\pm)$ and the $3\times 3$ matrices of 
integral operators $K(E)$ can be extracted from the expressions 
in the previous section. 

To convert $K(E)$ into a matrix we use Gauss-Legendre quadrature
to replace the momenta $p$ and $q$ by discrete variables.
We have examined three versions of this discretisation, in terms of
variables $x$ defined by: (1) $p = m \tan(x)$ where $m=140$~MeV, 
(2) $p=x$ (no change of variable), and (3) $p=m_0\exp(x)$ where
$m_0=1$~MeV. The variables
$p$ and $q$ run from zero to the cutoff $\Lambda$, except in the third 
of these where we also need to impose a cutoff at the very small 
momentum $m_0$.

We the construct the determinant of the matrix $K(E)-1$ and 
search on $E$ to find the energy eigenvalues at which 
$\det[K(E)-1]=0$. For one of these values of $E$, we then solve the
eigenvalue problem Eq.~(\ref{eq;Xx}) and determine the eigenvector
corresponding to the unit eigenvalue of the matrix $K(E)$.

For given values of the two-body input parameters we first set
the three-body force $H(\Lambda)$ to zero and adjust the value 
of the cutoff until the shallowest bound-state eigenvalue 
reproduces the observed three-body binding energy. Starting from 
this cutoff, $\Lambda_0$, we then determine the values of 
$H(\Lambda)$ needed to reproduce this binding energy for other
cutoffs.

\subsection{Triton}

In this case, we work with the determinant constructed from the
homogeneous part of the coupled integral equations in
Eqs.~(\ref{eq;a-iso-b}), (\ref{eq;b0-iso-b}), and (\ref{eq;bm-iso-b}).
The two-body input is provided by
\bea
\gamma = 45.703 \ \mbox{\rm MeV}\,,
\qquad 
a_{np} = -23.749 \pm 0.008 \ \mbox{fm \cite{kn-zpa75}}\,, 
\qquad
a_{nn} = -18.5\pm 0.4 \ \mbox{fm \cite{tg-prc87}}\,. 
\eea
We adjust the three-body force to fit the 
triton binding energy,
\bea
B_3(^3\mbox{\rm H}) = 8.48182 \ \mbox{\rm MeV}\,,
\eea 
which corresponds to a momentum scale $\alpha = \sqrt{m_N\,B_3} =89.239$~MeV.

For 16 Gaussian integration points (corresponding to a 48$\times$48
matrix) we find that our results have converged and, for the 
three choices of variable discussed above, they agree to 6 significant 
figures. 

\begin{figure}[h]
\includegraphics[width=14cm,keepaspectratio,clip]{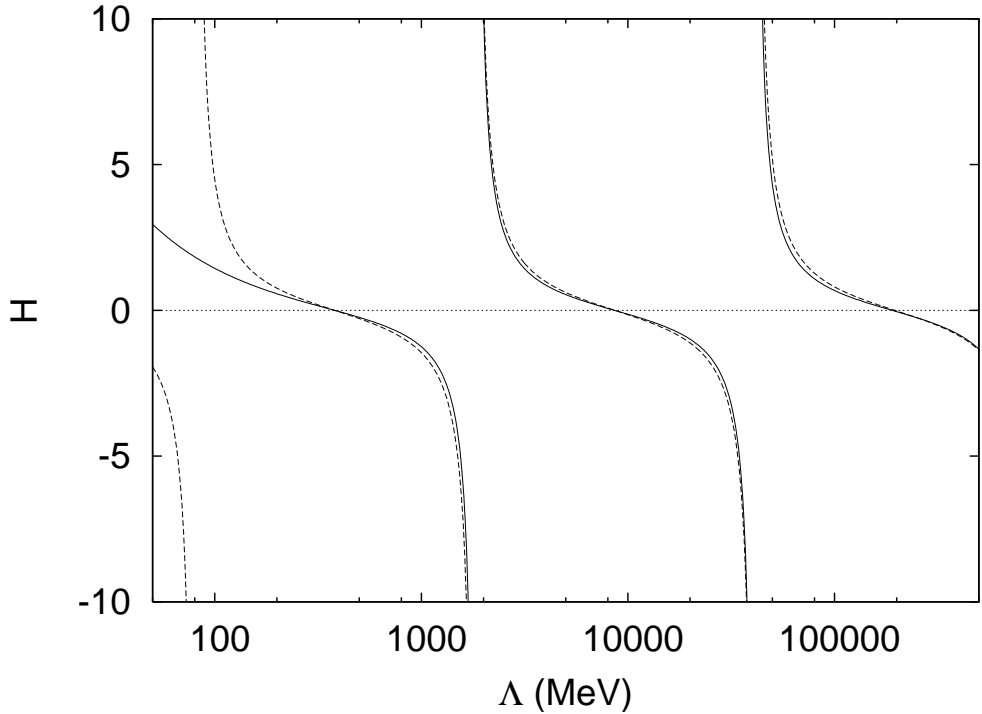}
\caption{Values of the three-body force $H$ as a function of the 
cutoff $\Lambda$ in MeV. For comparison, the dashed curve shows 
the asymptotic, scale-free form of $H(\Lambda)$.     }
\label{fig;H-triton-isb}
\end{figure}

For $H(\Lambda)=0$, we reproduce the experimental binding energy
with the cutoff $\Lambda_0=380.689$~MeV. The values of the three-body 
force needed for other cutoffs are shown in Fig.~\ref{fig;H-triton-isb}.

As we take $\Lambda\to \infty$, the cutoff becomes much larger than 
the physical scales in the system ($1/a$) and $H(\Lambda)$ displays
the limit-cycle behaviour found in Ref.~\cite{bhk-npa00}.
This reflects the presence of the Efimov effect in this three-body
system, a tower of deeply bound states with energies in a constant
ratio \cite{e-sjnp71}. In three-nucleon systems only the shallowest
of these states lies within the domain of the EFT and so corresponds to 
a physical state. 

The asymptotic form of $H(\Lambda)$ in this limit is \cite{bhk-npa00}
\be
H(\Lambda) = 
- \frac{\sin[s_0\ln(\Lambda/\Lambda_*)-\arctan(1/s_0)]}{
\sin[s_0\ln(\Lambda/\Lambda_*)+\arctan(1/s_0)]}\,.
\label{eq;H}
\ee
where $s_0=1.00624\cdots$ and the scale parameter $\Lambda_*$ is
defined by
\be
\Lambda_0 = \Lambda_*\exp[(1/s_0)\arctan(1/s_0)]\,.
\ee
This form is shown by the dashed line in Fig.~\ref{fig;H-triton-isb}.
The numerical results deviate from it only for small values of $\Lambda$ 
where the finite scattering lengths can influence the renormalisation
of the three-body force.

\subsection{${}^3$He channel}

For the ${}^3$He channel, we need to the $pp$ Coulomb scattering length 
as a two-body input parameter,
\be
a_C=-7.8063\pm 0.0026 \ \mbox{fm \cite{bcrs-prc88}}.
\ee
The experimental binding energy of ${}^3$He is
\be
B_3({}^3\mbox{\rm He}) = 7.71804\ \mbox{\rm MeV}\,,
\ee
corresponding to a scale $\alpha=85.127$~MeV

As already noted, the $K^{(d,V')}$ term contains a logarithmic
singularity that needs to be dealt with carefully. We handle it 
with the method outlined in Appendix B in Ref.~\cite{ksh-75}.
In the case of a bound state, the other factors in the integral
are regular and so the logarithmic singularity has a finite integral.
We can take the interval around the singular point and,
treating the other factors as constant, explicitly integrate 
the logarithm over it. The result can be expressed in a similar
form to a Gaussian quadrature,
\be
\int^b_a dq\, \ln(p_j-q)^2\, y(q) = \sum_{i\neq j}^N w_i \,
\ln(p_j-q_i)^2\, y(q_i)+w_j^\prime\, y(p_j)\,,
\ee
where $y(q)$ is a smooth function and the $w_i$ are the usual
weights. The modified weight for the singular point is given by
\bea
w_j^\prime &=& \int^{q_{j>}}_{q_{j<}}dq\, \ln(p_j-q)^2\nnb\\
&=&
-2w_j+2(q_{j>}-p_j)\ln(q_{j>}-p_j)
+2(p_j-q_{j<})\ln(p_j-q_{j<})\,,
\eea
where $q_{j<}=a+\sum_{k=1}^{j-1}w_k$ and $q_{j>}=a+\sum_{k=1}^jw_k$.

To examine the size of Coulomb effects in the three-nucleon system, 
we first consider an isospin-symmetric three-body force and take
$H(\Lambda_0)$ for the same value $\Lambda_0=380.689$~MeV as in the 
triton. We use $N_4=32$ Gaussian points for the four integrals needed
to evaluate the kernels $K^{(c)}$ and $K^{(d)}$. The integral
equations are discretised using up to $N=32$ points 
(leading to $96\times 96$ matrices). We fit the results for $N=16$ 
and 32 to the form $\alpha(N) \simeq A + B/N$ in order to extrapolate 
to $N=\infty$. 

Using the first choice for the discretisation variable ($p = m \tan(x)$), 
we get $\alpha=84.8084$~MeV, or a binding energy of $B_3=7.66038$~MeV.
This differs from the triton energy for the same three-body force
by 0.82~MeV. It is within 1\% of the observed $^3$He energy,
indicating that the isospin-violating three-body force is indeed
a higher-order contribution. This is as expected, given the absence of
any modification of the renormalisation, since the isospin-violating 
term is suppressed by one power of the inverse Bohr radius $\kappa$ 
relative to the LO force.

\begin{figure}[h]
\includegraphics[width=14cm,keepaspectratio,clip]{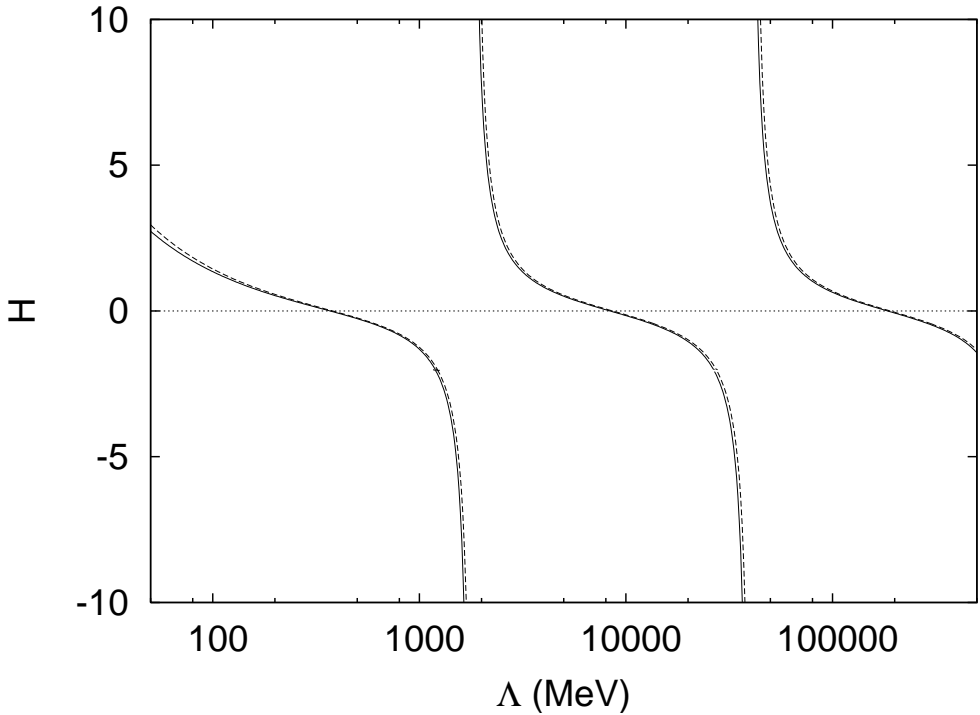}
\caption{Evolution with $\Lambda$ of the three-body forces  
$H(\Lambda)$ for $^3$He (solid line) and the triton (dashed).}
\label{fig;H-3he}
\end{figure}

Our difference between the triton and $^3$He binding energies 
is significantly larger than the $0.66\pm0.03$~MeV found by
Kirscher \textit{et al.}~\cite{kgsh-epja10}. However, although
working to NLO, those authors took a charge-independent value 
for the $^1S_0$ scattering length. Their results are consistent
with other estimates of the pure Coulomb contribution 
\cite{cb-prc87,wis-prl90}. 

Because we take different values for the $nn$ and $np$ $^1S_0$ 
scattering lengths, our calculations do include some higher-order 
isospin-breaking effects. Given that we have not done a consistent 
calculation to N$^2$LO (or even NLO), the best we can say is that 
the $\sim$160~keV difference between our result and that of 
Ref.~\cite{kgsh-epja10} is comparable in size to the contributions 
of 70--110~keV that have been estimated using other approaches 
\cite{cb-prc87,wis-prl90,mm-prc01,fpvk-prc05}. One caveat that 
should be made about these comparisons is that, as noted above, 
there is no model-independent way to separate Coulomb and other 
isospin-breaking effects in this EFT.

By adjusting the strength of the three-body force, we can fit 
the experimental binding energy. As in the triton case, we have done
this for a range of cutoffs. The results are plotted in 
Fig.~\ref{fig;H-3he}. They display the same limit-cycle behaviour
as in the triton channel. Moreover the differences between the 
forces needed in the two channels are very small, reflecting the 
fact that a symmetric force can give a very good account of $^3$He.

To quantify the size of the isospin breaking in the three-body forces,
it is better not to consider the values of $H(\Lambda)$, since these
cycle between $\pm\infty$. The scale $\Lambda_0$ or, equivalently 
$\Lambda_*$, is a more appropriate measure. Its value of 
$\Lambda_0=382.46$~MeV differs from that for the triton by 0.46\%.
The contribution of this force to the triton-$^3$He splitting is about
50~keV. This is much larger than the 5~keV estimate of the three-body 
contribution in Ref.~\cite{fpvk-prc05}, which may reflect the fact that
other isospin-breaking effects are missing from our calculatation. Also,
the results of Ref.~\cite{fpvk-prc05} were obtained using an EFT that
includes pion fields.

\section{Summary}

We have studied $^3$He within the framework of the pionless EFT,
treating the Coulomb interaction nonperturbatively. We solve
the set of integral equations developed by Kok 
\textit{et al.}~\cite{ksh-75}, which can be thought of as extending 
the equation of Skornyakov and ter-Martirosian \cite{stm-jetp57} to 
include the full off-shell Coulomb $T$-matrix. 

We find that a three-body interaction is required at leading order 
in $^3$He, as in the triton. This force also exhibits the same 
limit-cycle behavior as a function of the cutoff $\Lambda$ as found
by Bedaque \textit{et al.}~\cite{bhk-npa00} for systems with purely 
short-range interactions. These results show that the Coulomb interaction 
is not singular enough to alter the deeply bound ``Efimov" states in the 
three-nucleon system.

We find that an isospin symmetric three-body force, fit to the triton
binding energy, can give the $^3$He binding to better than 1\%. The 
scale parameter of the force that fits $^3$He differs from the corresponding 
value for the triton by about 0.5\%. This is of the expected size for an 
order-$\alpha$ electromagnetic effect.

Our results demonstrate that the Coulomb interaction has no nonperturbative
effect on the renormalisation of the three-body force in $^3$He. Moreover the 
overall contributions of the Coulomb interaction to the binding energy of 
$^3$He are small, perhaps unsurprisingly given the typical momenta involved, 
and so could have been calculated perturbatively. This will not be the case 
if our approach is extended to describe low-energy $pd$ scattering. However, 
as noted above, such an extension will require improved numerical techniques 
to handle the singularity of the Coulomb interaction. 

Our approach could also be used to determine the $^3$He wave function from the
corresponding eigenvector of the set of integral equations. This could then be 
used in calculations of electromagnetic properties of $^3$He, based on the 
methods in Ref.~\cite{sb-npa05}. 

Our present results do include certain other isospin-breaking effects beyond 
LO in this EFT, through the $nn$ and $np$ $^1S_0$ scattering lengths. A proper 
treatment of these in the three-nucleon system will require a consistent 
higher-order calculation. We have provided here the basic forms of the integral 
equations allowing for isospin-breaking in the NN scattering lengths. The 
relevant power counting for the terms needed in a full calculation can be 
determined using the techniques in Refs.~\cite{bghr-npa03,bb-jpa05}.

\section*{Acknowledgments}

This work was supported by STFC grants PP/F000448/1 and ST/F012047/1.
We are grateful to L. Kok and his colleagues in Groningen for providing 
copies of the reports \cite{ksh-75}.

\end{document}